# Rapid Microwave-Assisted Synthesis of Dextran-Coated Iron Oxide Nanoparticles for Magnetic Resonance Imaging

Elizabeth A. Osborne[1], Tonya M. Atkins[1], Dustin A. Gilbert[2], Susan M. Kauzlarich[1], Kai Liu[2], and Angelique Y. Louie[3*]

[1]Department of Chemistry, [2]Department of Physics, and [3]Department of Biomedical Engineering, University of California, Davis, California 95616, USA

*aylouie@ucdavis.edu

## ABSTRACT

Currently, magnetic iron oxide nanoparticles are the only nano-sized magnetic resonance imaging (MRI) contrast agents approved for clinical use, yet commercial manufacturing of these agents has been limited or discontinued. Though there is still widespread demand for these particles both for clinical use and research, they are difficult to obtain commercially, and complicated syntheses make in-house preparation infeasible for most biological research labs or clinics. To make commercial production viable and increase accessibility of these products, it is crucial to develop simple, rapid, and reproducible preparations of biocompatible iron oxide nanoparticles. Here, we report a rapid, straightforward microwave-assisted synthesis of superparamagnetic dextran-coated iron oxide nanoparticles. The nanoparticles were produced in two hydrodynamic sizes with differing core morphologies by varying the synthetic method as either a two-step or single step process. A striking benefit of these methods is the ability to obtain swift and consistent results without the necessity for air, pH, or temperature sensitive techniques; therefore, reaction times and complex manufacturing processes are greatly reduced as compared to conventional synthetic methods. This is a great benefit for cost-effective translation to commercial production. The nanoparticles are found to be superparamagnetic and exhibit properties consistent for use in MRI. In addition, the dextran coating imparts the water-solubility and biocompatibility necessary for *in vivo* utilization.

Iron oxide particles have been widely studied for commercial applications in magnetic storage media, catalysis, magnetic inks, pigments, ferrofluid seals, biotechnology, and medical diagnostics.[1] In particular, advances in the fabrication of nanosized magnetic particles has led to novel and rapidly evolving uses in biomedical applications such as cell tracking and separation, hyperthermia treatment, immunoassays, drug delivery vehicles, and magnetic resonance imaging (MRI) contrast agents.[1-3] For such applications *in vivo*, the ideal iron oxide nanoparticles should exhibit a small size (< 100 nm) with minimal polydispersity, a nontoxic, biocompatible surface coating, and high magnetization in the presence of a magnetic field. Development of superparamagnetic iron oxide nanoparticles (SPIO) with a variety of differing cores and surface coatings for molecular imaging has resulted in numerous contrast agents that are currently in commercial use or under clinical investigation.[2,4] Yet, consistency of successful preparations that are time-, energy-, and cost-efficient is always of concern.

The scientific literature presents a plethora of techniques to prepare magnetic iron oxide nanoparticles including coprecipitation,[5-15] thermal decomposition,[16-19] sonolysis,[20-24] electrochemical deposition,[25] sol-gel processes,[26-28] spray and laser pyrolysis,[29-31] flow injection synthesis,[32] hydrothermal and high temperature syntheses,[33-38] and nanoreactors such as protein cages,[39-42] vesicles,[43,44] and microemulsions[45-48]. While each of these methods may enjoy certain advantages depending on the desired properties of the iron oxide nanoparticles, many of them possess severe disadvantages for clinical translation such as the necessity for costly, specialized equipment, low yielding, excessively complicated synthetic schemes, and uncontrollable or inconsistent results. An example of a circumstance where manufacturing process vs. cost was critical, is in the case of Feridex®, the only U.S. FDA approved SPIO

MRI contrast agent. Feridex® was approved by the FDA in 1996 and likewise, obtained approval in overseas markets as well. Despite worldwide distribution and use of this SPIO contrast agent, Feridex® is no longer being marketed by its proprietary owner and lone manufacturer, AMAG Pharmaceutical, Inc. The company announced on their website that production of the contrast agent was halted in November 2008. Though a reason was not specified, one can infer that the company's net profits from Feridex® were not sufficient to continue commercial manufacture of the agent. Perhaps more time- and cost-efficient production would have changed this outcome.

By far, the most common and classically used method for synthesis of SPIO is coprecipitation of $Fe^{3+}$ and $Fe^{2+}$ salts in a basic medium.[4] The coprecipitation method offers some advantages in that a large quantity of particles may be prepared and the size, shape, and composition of the nanoparticles can be modified depending upon the ratio and type of iron salts used, the reaction pH, and the ionic strength of the medium.[49-53] In addition, researchers have demonstrated the ability to apply surface coatings to the particles both during and following the coprecipitation process. Nonetheless, a major drawback of this method is limited control of size distribution.[4,54,55] Despite the ease and efficiency of coprecipitation, extensive purification or size sorting may be necessary.[2,56] For surfactant stabilized particles, Perales-Perez and colleagues used physicochemical methods to effectively select 4-7nm magnetite particles from polydisperse powder with sizes ranging from 4-40nm.[57] For polymer coated particles, Rheinlander et al. demonstrated using flow field-flow fractionation to sort a 10-50nm mixture of dextran-coated iron oxide particles into approximately 8 different batches according to size.[58] Magnetic properties vary depending on nanoparticle size; therefore, the ability to control the dispersity of a sample is crucial.[59] Several methods have been proposed for particle size sorting in recent years, including common physical processes such as ultracentrifugation, microfiltration, or size exclusion chromatography.[57,60] Also, size sorting may be performed by adding an electrolyte solution to the particle solution to disrupt its stability and cause precipitation of large particles.[61] Less common methods such as magnetic chromatography[56] have also been explored. Each of these methods lengthens and complicates the synthetic process and reduces the yield of the desired product. As a result, researchers are seeking new means that offer even greater advantages without the inherent shortcomings of the current methodology. A potential answer lies in the emergent microwave synthesis technology.

Microwave-assisted synthesis has been utilized since the late 1980s in the preparation of organics, organometallics, and peptides.[62] Just in the last decade, exploration of this methodology for inorganic material synthesis has been gaining momentum. Microwave synthesis has been shown to significantly reduce reaction time, increase yields, reduce side reactions, enhance reproducibility, and provide a more energy efficient, greener process.[62,63] Microwave heating presents significant benefits over traditional heating methods (i.e. oil bath), which rely on conduction and convection for heat distribution. Microwave radiation heats materials through much more efficient dielectric heating; this occurs as molecular dipoles attempt to align with the alternating electric field. Thus, the heating phenomenon depends on a substance's ability to absorb microwaves and convert the energy to heat; generally, more polar solvents, reagents, and catalysts are more efficiently heated. Beyond the previous noted benefits, in the case of nanoparticle synthesis, microwave preparation has yielded greater control of size and dispersity as well as enhanced crystallinity.[64]

Microwave syntheses for superparamagnetic iron oxide nanoparticles have been reported that yield $Fe_2O_3$ particles[65-68], mixed $Fe_2O_3$ and $Fe_3O_4$ particles[69], and $Fe_3O_4$ particles[67,70-74]. Only a small fraction of those reported indicate passivation of the nanoparticle surface, which is necessary for water solubility and biocompatibility. In 2007, Zhu and coworkers reported microwave synthesis of irregular shaped $Fe_3O_4$ nanoparticles and ellipsoidal $Fe_2O_3$ nanoparticles that were coated with polyethylene glycol (PEG-20000).[67] In 2009, Yang et al published the microwave synthesis of spherical, nanoporous $Fe_3O_4$ nanoparticles also coated with PEG.[71] In 2010, Edrissi and colleague reported the microwave-assisted synthesis of spheroid $Fe_2O_3$ nanoparticles coated with linoleic or palmitoleic acid.[74] To date, no microwave methods to prepare dextran-coated iron oxide nanoparticles have been reported.

Dextran is an abundant, inexpensive polymer that consists solely of alpha-D-glucopyranosyl monomers. The polysaccharide has a large number of hydroxyl groups that foster chelation and hydrogen-bonding interactions with iron oxide surfaces as well as provide locations for modification.[2] Dextran coating provides stability and biological compatibility to the iron oxide nanoparticles, thereby making them suitable MRI contrast agents.[4] Due to their superior biocompatibility properties, this platform is a common motif for MR T2 agents. Previous pharmacokinetic studies have established that dextran-coated iron oxide nanoparticle are sequestered by the reticuloendothelial system and metabolized to molecular iron for incorporation into hemoglobin within 5-40 days.[75,76] Herein, we report the synthesis of superparamagnetic dextran-coated iron oxide nanoparticles by two separate microwave heating methods. In varying the synthetic strategy by either a two-step or single step process, we were able to consistently obtain iron oxide nanoparticles in two size regimes. The resulting nanoparticles were characterized with transmission electron microscopy (TEM), dynamic light scattering (DLS), powder X-ray diffraction (XRD), Fourier transform infrared spectroscopy (FT-IR), inductively coupled plasma mass spectrometry (ICP-MS), vibrating sample magnetometry (VSM), superconducting quantum interference device (SQUID) magnetometry and magnetic resonance relaxometry.

**RESULTS AND DISCUSSION**

Microwave-assisted synthesis of dextran-coated iron oxide nanoparticles (DIO) was successfully accomplished by two methods. In the first approach, the nanoparticles were synthesized via a two-step reaction. First, uncoated iron oxide nanoparticles, basically iron cores with no coating, were prepared by hydrazine reduction of ferric chloride with microwave heating at 100°C for 10 minutes. Subsequent coating of the nanoparticles with dextran was achieved in a second stage of microwave heating at 100°C for 2 minutes in the presence of additional ferric chloride, sodium hydroxide, and reduced dextran. The coating step was more challenging than first anticipated, as numerous attempts were unsuccessful. The initial trials, comprising reactions of either reduced dextran with bare nanoparticles or reduced dextran with bare nanoparticles in acidic or basic conditions, resulted in large aggregates of particles that quickly fell out of solution. Successful dextran coating was finally achieved by inclusion of an additional 400 μL of 1.0M ferric chloride in the second coating step. Presumably, the $Fe^{3+}$ ions associate with the nanoparticle surface and function to provide a charged particle shell, thus sustaining the particle dispersion in aqueous solution and encouraging dextran passivation.[77] In the second synthetic approach, the nanoparticles were synthesized in a one-pot single step microwave reaction. Ferric chloride and reduced dextran were reacted with hydrazine in the microwave at 100°C for 10 minutes. The two methods result in different sized nanoparticles.

TEM was used to characterize the size and shape of the iron oxide cores. TEM images in Figure 1 demonstrate the differing morphology obtained with the different synthetic schemes. The single step synthesis resulted in relatively monodisperse particles with an iron core size of 6.5 ± 1.2 nm. In contrast, two-step synthesis results in larger size particles with the iron oxide cores measuring 18.0 ± 4.1 nm in a mixture of shapes ranging from spheroid to cubes. Predictably, the dextran coating step did not change the observed core size as uncoated particles after step 1 similarly measured 17.7 ± 6.6 nm (Fig. 1b). However, the shape has changed and the large square looking particles are noticeably missing from the TEM image. The histograms beneath each image in Figure 1 provide a look at the entire population distribution of the nanoparticle core diameters used to calculate the mean size ± one standard deviation. Dynamic light scattering (DLS) was used to determine the hydrodynamic size of the particles in aqueous solution. While the dextran coating is translucent to TEM, the DLS measurement provides the size of the particles with coating included. The two-step synthesis produced nanoparticles with an overall hydrated diameter of 67 ± 17 nm, while the single step method yielded 39 ± 8 nm particles. These DLS results are consistent with values previously obtained in our lab for dextran coating thickness and iron core size with conventional heating methods.[78-80] Therefore, compared with synthetic methods that yield wide size ranges, there is promising efficiency for production by the microwave method. In this case, on average, 85% of the one-step product is between 30-50 nm and 62% of the two-step product is 50-80 nm in size. In addition, the reduced reaction time and reduced complexity of the technique make it an attractive alternative.

Furthermore, the smaller iron core size for nanoparticles formed while dextran is *in situ* versus subsequent addition has been reported in literature.[81-83] Dextran is thought to limit the growth of the iron oxide core by confining the space available for crystal growth. At high dextran concentration, the randomly folded polymer provides small spaces for nanoparticle nucleation and growth that is halted upon dextran adsorption to the iron oxide surface.[82,83]

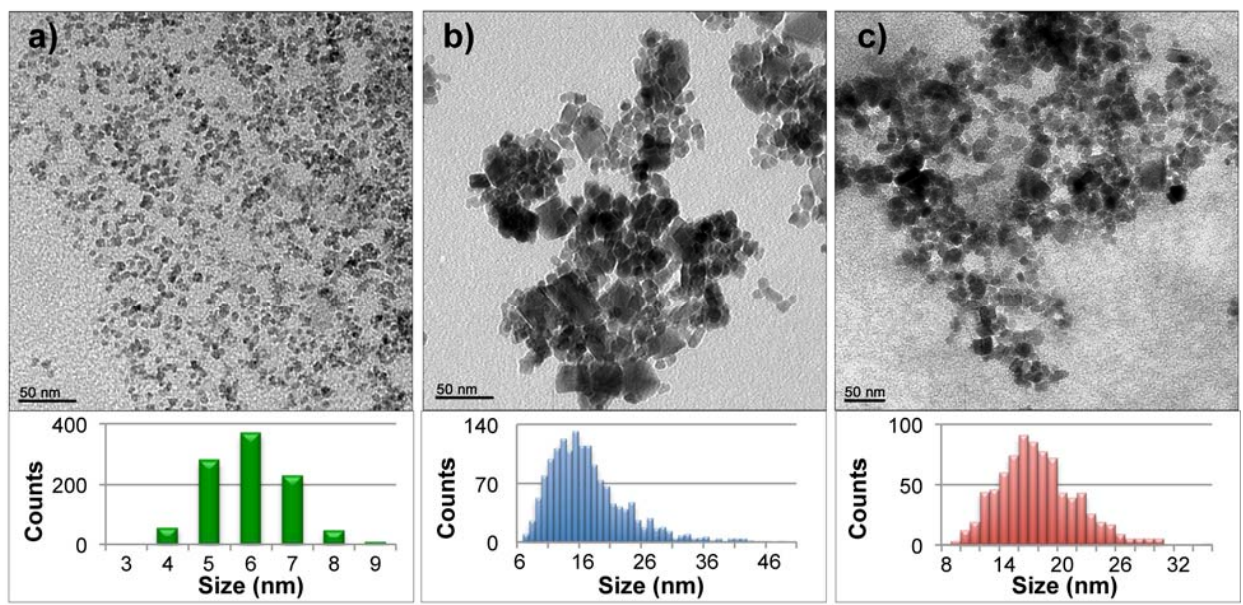

**Figure 1. Representative TEM images of (a) one-pot synthesized dextran-coated iron oxide nanoparticles (DIO) with an iron oxide core measuring 6.5 ± 1.2 nm, (b) uncoated iron oxide nanoparticles (obtained after step 1 of the two-step synthesis) with a core diameter of 17.7 ± 6.6 nm, and (c) two-step DIO with a core diameter of 18.0 ± 4.1 nm. Scale bar = 50 nm. The histograms included below each image indicate the population size distributions for three batches of each, respectively.**

The crystal structures of representative samples of the dextran-coated and uncoated nanoparticles were analyzed by powder X-ray diffraction (XRD). Though line broadening slightly obscures the precision of the diffraction patterns, the line position and intensity features (Fig. 2) were consistent with the spinel phases of magnetite ($Fe_3O_4$) and/or maghemite ($\gamma$-$Fe_2O_3$). These two phases are not distinguishable from the obtained XRD patterns; therefore, we suggest that the product may contain either or both of these iron oxide phases. The reference peaks for magnetite and maghemite are provided (in Fig. 2d/e) for comparison to the prepared nanoparticles. The line broadening observed for XRD patterns of nanosized crystalline particles results from diffraction of X-rays over volumes with size analogous to the

X-rays wavelength.[84] This effect allows for calculation of the crystallite size using the Scherrer equation assuming one product and a Gaussian size distribution. Calculation of the crystallite sizes yielded 4.3 nm ± 0.2 (whole pattern fitting assuming all $Fe_3O_4$) for the one-step iron oxide, and this is slightly smaller than that determined by TEM. The powder diffraction patterns of the uncoated iron oxide nanoparticles and the two-step DIO product show reflections that are composed of a broad base indicative of small particles superimposed with a narrow peak indicative of much larger particles, consistent with the large size distribution observed in the TEM. This is especially apparent for the powder diffraction of the two-step DIO. Whole pattern fitting is only appropriate for obtaining the average crystallite size when the samples are fairly uniform. While the average crystallite size is consistent with the TEM for the one-step method, in the case of the uncoated particles and two-step DIO, crystallite size from powder X-ray diffraction would be misleading because of the very large size distribution observed in the TEM.

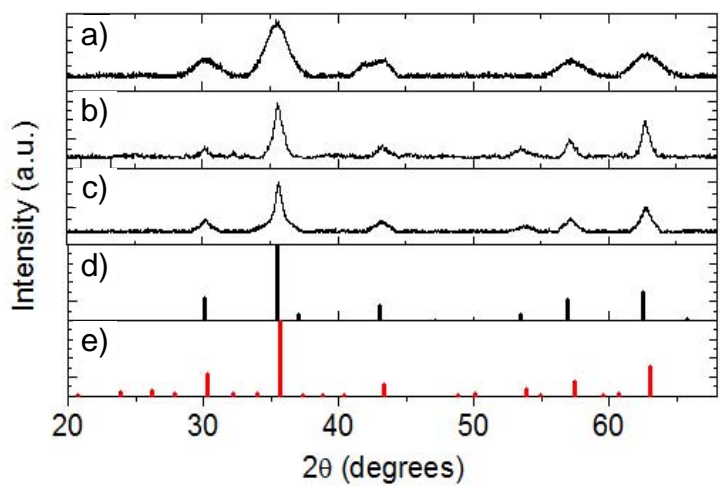

**Figure 2.** XRD patterns for (a) one-step DIO, (b) uncoated iron oxide particles (obtained after step 1 of the two-step synthesis), (c) two-step DIO, and reference patterns for (d) $Fe_3O_4$ PDF #65-3107 and (e) $\gamma$-$Fe_2O_3$ PDF # 25-1402.

Percent iron content for the DIO nanoparticles was determined by ICP-MS in order to allow calculation of magnetic properties per mass of iron. Given the percent of iron per mass of sample, the magnetization of the particles was evaluated by VSM and plotted with respect to the amount of iron (emu/gram of Fe). Room temperature magnetic hysteresis loops of dextran-coated iron oxides are shown in Fig. 3. For the one-step synthesis, the sample is superparamagnetic with no hysteresis (Fig. 3a). On the contrary, a small ferromagnetic component was observed for the two-step product (Fig. 3b), consistent with the larger iron oxide core observed by TEM. Notably, small but appreciable remanent magnetization only appeared after the dextran coating step for the two-step synthesis; the magnetization curve for the uncoated nanoparticles after step 1 of the two-step synthesis (not shown) was similar to that of the one-step product with no hysteresis. The dextran coating step likely caused further grain growth, albeit small, that pushed the superparamagnetic blocking temperature above room temperature.

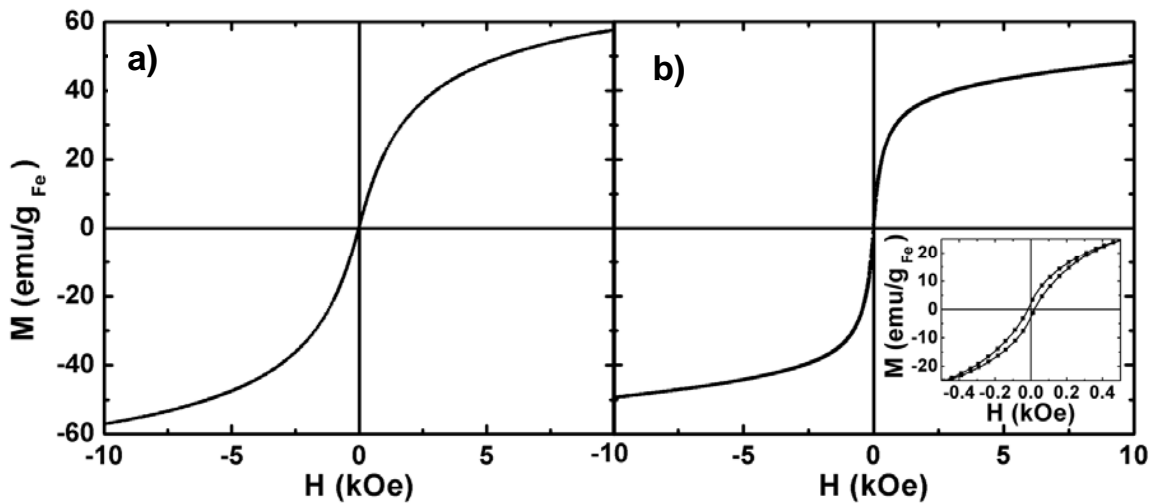

**Figure 3. Room temperature magnetic hysteresis loops of (a) one-step iron oxide nanoparticles and (b) two-step iron oxide nanoparticles with an inset showing the small hysteresis loop, indicating a ferromagnetic component.**

The temperature dependence of the magnetization was analyzed by SQUID magnetometry. For the one-step DIO sample, the zero-field-cooled (ZFC) curve displays a maximum in the temperature dependence of the magnetization, corresponding to a blocking temperature of ~25 K (Fig. 4a). In comparison, the two-step DIO sample is ferromagnetic at room temperature. Saturation magnetization at 10 K for the one-step and two-step sample is 71 emu/$g_{Fe}$ (Fig. 4b) and 42 emu/$g_{Fe}$, respectively. Following the method laid out in Cho et al.,[85,86] from the blocking temperature, we estimate the size of the one-step particles to be 5.7 nm, consistent with the TEM imaging. Similarly, the size of the two-step particles is >13 nm, again consistent with TEM measurements. Despite the larger iron oxide core, the two-step microwave prepared particles exhibited lower magnetization suggesting the potential presence of phases that are less magnetic. In literature, reduced magnetization due to surface characteristics and coatings has also been observed and attributed to several mechanisms, including spin-canting at the particle surface or a surface layer that is magnetically ineffective.[3] Further examination beyond the scope of this paper is required to illuminate the effect of the nanoparticle surface coating on magnetization.

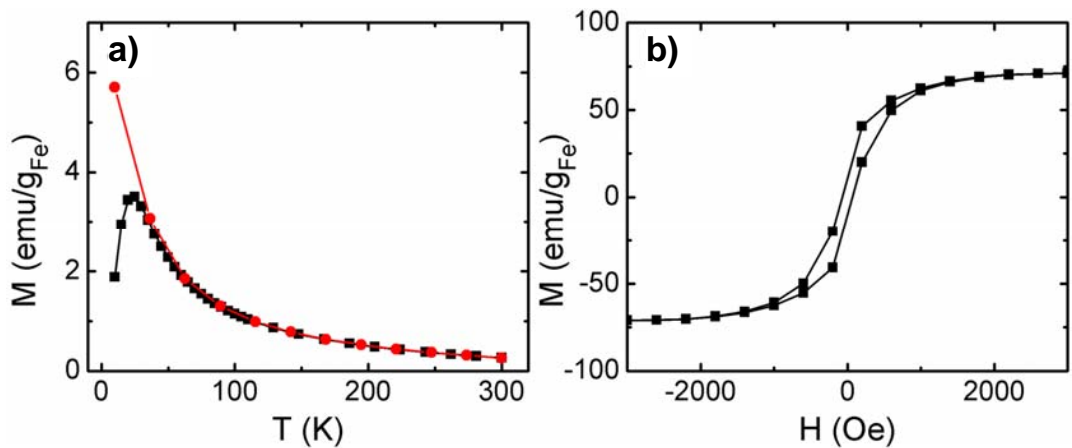

**Figure 4. Temperature dependence of magnetization of the one-step dextran-coated iron oxide nanoparticles measured in a 10 Oe external field, (a) ZFC (black squares) and FC (red circles) curves and (b) magnetic hysteresis loop at 10 K, normalized to the mass % of Fe.**

Infrared spectroscopy (IR) confirmed the presence of dextran coating as revealed by comparison of both uncoated and dextran coated iron oxide nanoparticles (Figure 5). The uncoated nanoparticles (red), obtained after step 1 of the two-step synthesis, display the Fe-O vibration band at 586cm$^{-1}$. Following the addition of dextran, characteristic absorption bands for hydroxyl at 3387cm$^{-1}$ (broad), C-O at 1018cm$^{-1}$, and C-H at 2932cm$^{-1}$ were observed for both the one- (blue) and two-step (green) dextran coated nanoparticles, while the Fe-O band was slightly obscured. In concurrence with the measured hydrodynamic size, these results further verified the coating of dextran onto the iron oxide nanoparticles.

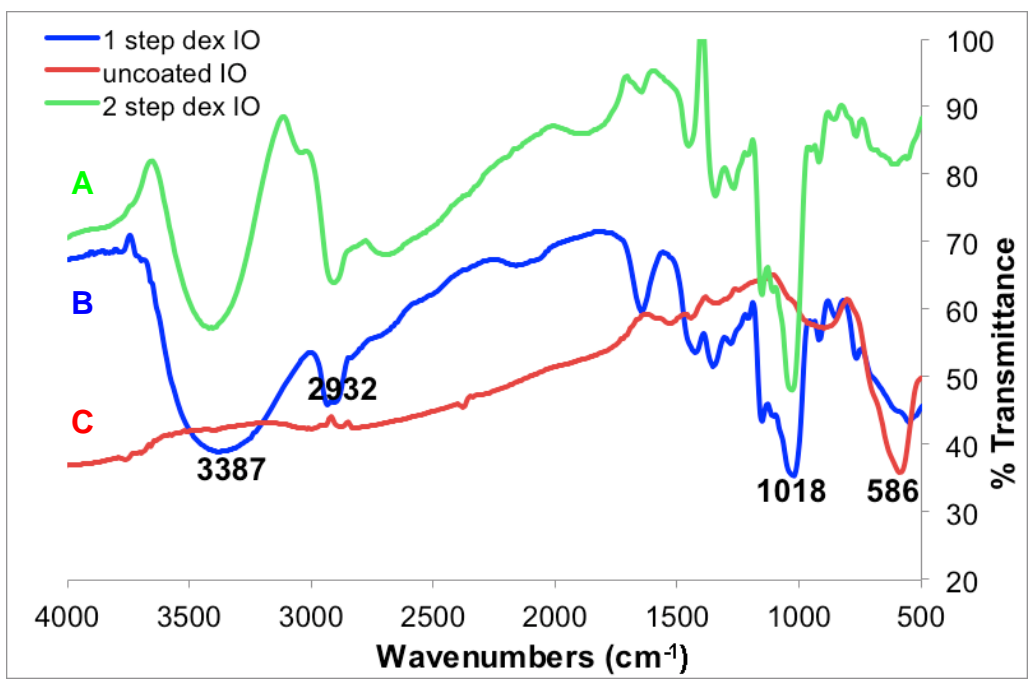

**Figure 5.** IR spectra of one-step iron oxide nanoparticles (B, blue), two-step iron oxide nanoparticles (A, green), and uncoated IO particles (C, red).

Relaxivity, a concentration-independent measure of the effectiveness of a paramagnetic material, was determined as the slope of inverse relaxation time versus iron concentration. Superparamagnetic iron oxide nanoparticles are commonly considered $T_2$ contrast agents due to their high transverse relaxivities.[2] Longitudinal and transverse relaxation time of aqueous iron oxide nanoparticle samples were measured at 60 MHz and 37 °C. Table 1 includes the calculated relaxivities for the two iron oxide microwave preparations. The one-step product yielded an $r_2$ relaxivity of 58.1 ± 3.0 mM$^{-1}$ s$^{-1}$ while the two-step DIO was 39.3 ± 7.4 mM$^{-1}$ s$^{-1}$. While we initially expected the larger core nanoparticles to produce the greater relaxivity, these results are consistent with the magnetization data in which the two-step particles generate smaller magnetization. As a reference, the formerly commercially manufactured SPIO, Feridex®, was reported to have an $r_1$ of 10.1 mM$^{-1}$ s$^{-1}$ and an $r_2$ of 120 mM$^{-1}$ s$^{-1}$ at 1.5 T field strength.[2] Though the relaxivity of Feridex® is roughly twice that of our one-pot DIO, the wide size distribution of Feridex® (60-150nm) is not ideal for uniform biodistribution.

**Table 1. Longitudinal and Transverse Relaxivity of Dextran-coated Iron Oxide Nanoparticles**

|              | $r_1$ (mM$^{-1}$ s$^{-1}$) | $r_2$ (mM$^{-1}$ s$^{-1}$) |
|--------------|----------------------------|----------------------------|
| Two-step DIO | 2.3 ± 0.3                  | 39.3 ± 7.4                 |
| One-step DIO | 7.8 ± 0.3                  | 58.1 ± 3.0                 |

**CONCLUSIONS**

In conclusion, this work describes two methods for the microwave-assisted synthesis of superparamagnetic dextran coated iron oxide nanoparticles. These simple and rapid preparations allow for consistent synthesis of dextran-coated iron oxide nanoparticles without the necessity for techniques such as prior chilling or degassing of reagents, strict pH control, or use of Schlenk line. Therefore, reaction times and complex manufacturing processes are greatly reduced, and the majority of the product is in a desirable size range, as compared to conventional synthetic methods. This is of utmost importance for cost-effective translation to commercial production. Based on the results described herein, the single-step, one-pot microwave synthesis shows greater promise over the two-step method as a rapid and consistent method to produce iron oxide nanoparticles for MRI. The one-step method yields dextran passivated iron oxide nanoparticles that are fairly monodisperse and of favorable size (hydrodynamic diameter of ~39 nm and 85% of particles are 30-50 nm in size) to allow longer blood retention; also, its magnetization and relaxivity profiles are superior to that of the two-step product. In future work, we will determine what effect a change in synthetic parameters, such as stoichiometry or microwave reaction time and temperature, will have on the agent's morphology and MR characteristics.

Microwave synthesis of iron oxide nanoparticles offers exciting possibilities for rapid production of clinically relevant contrast agents and the ability to rapidly explore new platforms (i.e. Core-shell, doped, etc.) for future MR contrast agents with nearly instant results. In addition, in our previously published work, we have demonstrated the ability to modify dextran coatings to allow for further functionalization or targeting ability.[78-80,87] Thus, the potential to develop these nanoparticles, beyond blood-pool agents, into activatable MRI contrast agents is within reach.

## MATERIALS AND METHODS

General

All reagents were purchased from commercial sources and used without further purification. Nanopure water (18.0 MΩ·cm) from a Barnstead nanopure filtration unit was used throughout experiments. Microwave synthesis was carried out in a Discover SP/Explorer-12 Automated Microwave Synthesis System (CEM Corporation, USA) using glass microwave reaction vessels with Teflon caps purchased from CEM Corp. The temperature of each microwave reaction was precisely controlled by the Discover system; however, the reaction pressure was not controlled beyond setting the maximum at 250psi and allowing the vessels to self-vent as needed. FT-IR spectra were obtained using a Shimadzu IR Prestige 21 equipped with a diffuse reflectance accessory. Lyophilized nanoparticle powder was mixed with KBr, and the measurements were performed under ambient conditions.

Two-step synthesis of dextran-coated iron oxide particles (DIO)

a. *Step 1: Preparation of uncoated iron oxide nanoparticles.* In a typical microwave synthesis, ferric chloride ($FeCl_3 \cdot 6 H_2O$, 78 mg) was dissolved in 8 mL of water in a 35 mL microwave reaction vessel equipped with a stir bar. Immediately prior to microwave heating, hydrazine hydrate ($N_2H_4 \cdot H_2O$, 1 mL) was added to the vessel at room temperature with stirring.[67] The mixture was heated by microwave at 100 ± 5 °C for 10 min (300W max power, 250psi max pressure) with rapid stirring. The black product was collected by centrifugation, washed with water five times, and carried forward to step 2 or freeze-dried to yield a black crystalline powder.

b. *Step 2: Dextran coating of particles.* In preparation for dextran coating, the dextran (10,000 MW) was reduced with sodium borohydride according to a literature method.[88] Briefly, an aqueous solution of dextran was stirred with sodium borohydride (26 equivalents) at room temperature for 12 h. The reaction was stopped by the dropwise addition of conc. HCl until pH 7 and then the reduced dextran was dialysed

(Spectra/Por 6, MWCO 10,000) against nanopure water. The dextran was applied to freshly prepared uncoated iron oxide particles following the water washings. Reduced dextran (100 mg) was dissolved in an aqueous nanoparticle suspension (5 mL) by sonication. Then, 1.0 M ferric chloride (400 μL) and 1.0 M NaOH (1.0 mL) were added to the mixture preceding microwave heating at 100 °C for 2 min. The resulting solution was purified by membrane dialysis (Spectra/Por 6, MWCO 15,000) against nanopure water and freeze-dried to yield a brown crystalline solid.

Single step, one pot preparation of dextran-coated iron oxide particles (DIO)

Dextran coated iron oxide nanoparticles were synthesized in a single step by a microwave method similar to the uncoated nanoparticles described above. Ferric chloride (76 mg) and reduced dextran (100 mg) were dissolved in 8 mL of water with stirring in a 35 mL microwave reaction vessel. The ratio of dextran:Fe used (1:27) was chosen based on our previous experience in dextran coating preparation.[79,80] Immediately prior to microwave heating, hydrazine hydrate (1 mL) was added to the vessel at room temperature with stirring. The mixture was heated by microwave at 100 ± 5 °C for 10 min with rapid stirring. The product could not be collected by centrifugation due to increased stability in aqueous media. The black solution was purified by membrane dialysis (Spectra/Por 6, MWCO 15,000) against nanopure water and freeze-dried to yield a brown crystalline solid.

Characterization of size, iron content, and crystalline structure

Hydrated particle size was determined by dynamic light scattering (DLS) on a Nanotrac 150 particle size analyzer (Microtrac, Inc., Montgomeryville, PA). A geometric eight-root regression, with no residuals, was used to fit the data. The Nanotrac 150 has a built-in thermometer to measure the cell temperature, from which the viscosity is calculated; the nanorange option was enabled and scan time of three times 30 s was used. Particle size is expressed as the mean diameter ± one standard deviation. Transmission electron microscopy (TEM) images were utilized to determined iron oxide core size and shape (Phillips CM120, operating at 80kV and equipped with a Gatan Megascan digital camera). Sample solution was dropped onto carbon-coated 300 mesh copper grids. Core particle diameter was determined by taking the mean diameter of at least 300 particles for 3 separate syntheses as measured from TEM images in *Image J* software (National Institutes of Health). Iron content was determined by inductively coupled plasma mass spectrometry (ICP-MS) (Agilent Technologies 7500ce). The samples were prepared by digestion in 3% nitric acid. Powder X-ray diffraction (XRD) patterns were obtained on a Bruker D8 Advance X-ray diffractometer with Cu K$_\alpha$ radiation (1.5418 Å). The diffraction patterns were collected between 20° < 2θ < 68° with a Time/Step of 1.61 s and a step size of 0.006 (2θ). Data were analyzed with MDI Jade Plus 6.1.1 software. Average crystallite size was determined by the Whole Profile Fitting within the MDI Jade suite of programs for the freeze-dried product from the one-step synthesis. The other products were analyzed for crystallite size by the Scherrer equation for the (hkl) peak.

Magnetic Measurements

A Princeton Measurement Corp. MicroMag vibrating sample magnetometer (VSM) and a Quantum Design superconducting quantum interference device (SQUID) magnetometer were used to evaluate the magnetic properties of the iron oxide nanoparticle samples. The lyophilized nanoparticle powders were prepared by first measuring their mass, then placing each in a gel-cap with cotton stopper. The gel-cap was mounted on a quartz rod for insertion to the VSM. Each measurement was conducted in a field range of ±1.8T, with variable step size down to 2 Oe near remnance and a measurement time of 0.1 sec. Background measurements of the gel-cap plus stopper were conducted and found to be < 3% of the signal from the powders. A SQUID magnetometer was used to measure the zero-field-cooled (ZFC) and field-cooled (FC) temperature dependence of the magnetization in an external field of 10 Oe, as well as magnetic hysteresis loops at low temperatures (measurement time ~30s).

Relaxometry

Longitudinal $(T_1)$ and transverse $(T_2)$ relaxation times were measured on a Bruker Minispec mq60 relaxometer (Billerica, MA) at 60 MHz and 37 °C. $T_1$ values were measured using an inversion recovery sequence with 10–15 data points, and $T_2$ was measured using a Carr–Purcell–Meiboom–Gill (CPMG) sequence with $\tau$ = 1 ms, and 200 data points. Longitudinal and transverse relaxivity was determined as the slope of the line for plots of $1/T_1$ or $1/T_2$, respectively, against increasing iron concentration with a correlation coefficient greater than 0.99.

Acknowledgment

The authors wish to acknowledge funding from the Department of Energy (DESC0002289) and NSF (DMR-1008791) for support of this work.